# Ultrafast active plasmonics: transmission and control of femtosecond plasmon signals

K. F. MacDonald[1], Z. L. Sámson[1], M. I. Stockman[2], and N. I. Zheludev[1]*

[1] *Optoelectronics Research Centre, University of Southampton, Highfield, Southampton, Hampshire, SO17 1BJ, UK.*

[2] *Department of Physics and Astronomy, Georgia State University, 29 Peachtree Center Avenue, Atlanta, GA 30303-4106, USA.*

We report that femtosecond surface plasmon polariton pulses can propagate along a metal-dielectric waveguide and that they can be modulated on the femtosecond timescale by direct ultrafast optical excitation of the metal, thereby offering unprecedented terahertz plasmonic bandwidth - a key missing component in the development of surface plasmons as information carriers for next generation nanophotonic devices.

Surface plasmon polaritons (SPPs), propagating bound oscillations of electrons and light at a metal surface, have great potential as information carriers for next generation, highly integrated nanophotonic devices[1,2]. Since 2004, when the term 'active plasmonics' was coined in a paper reporting the concept of using optically-activated phase-change materials to control propagating SPPs[3], reversible changes in waveguide media caused by heating[4,5] or the modification of charge distribution[6], and interactions mediated by quantum dots[7], have been employed to control SPP signals. However, with sub-microsecond or nanosecond response times at best these techniques are too slow for current and future data processing architectures. Here we report that femtosecond plasmon pulses can propagate along a metal-dielectric waveguide and that they can be modulated on the femtosecond timescale by direct ultrafast optical excitation of the metal, thereby offering unprecedented terahertz plasmonic bandwidth - a speed at least five orders of magnitude faster than existing technologies.

In essence, we have discovered a nonlinear interaction between a propagating SPP and light that takes place in the skin layer of the metal surface along which the plasmon wave is propagating. A femtosecond light pulse incident on the metal surface disturbs the equilibria in the distributions of electron energies and momenta, thereby influencing plasmon propagation along the surface.

The nonlinear interaction between propagating SPP waves and light has been demonstrated in a pump-probe experiment wherein a pulsed plasmonic probe signal was generated on an Al/silica interface by grating coupling from a pulsed 780 nm laser beam. After travelling across the interface, the plasmon wave was decoupled to light by another grating and subsequently detected. Optical control (pump) pulses, originating from the same laser, were incident on the waveguide region between the coupling and de-coupling gratings (see Fig. 1). The transient effect of control pulse excitation on the propagation of the SPP signal was monitored by varying the time delay between the SPP excitation and optical pump pulses. It is found that an optical pump fluence of about 10 $mJ/cm^2$ leads to around 7.5% modulation of the plasmon wave intensity.

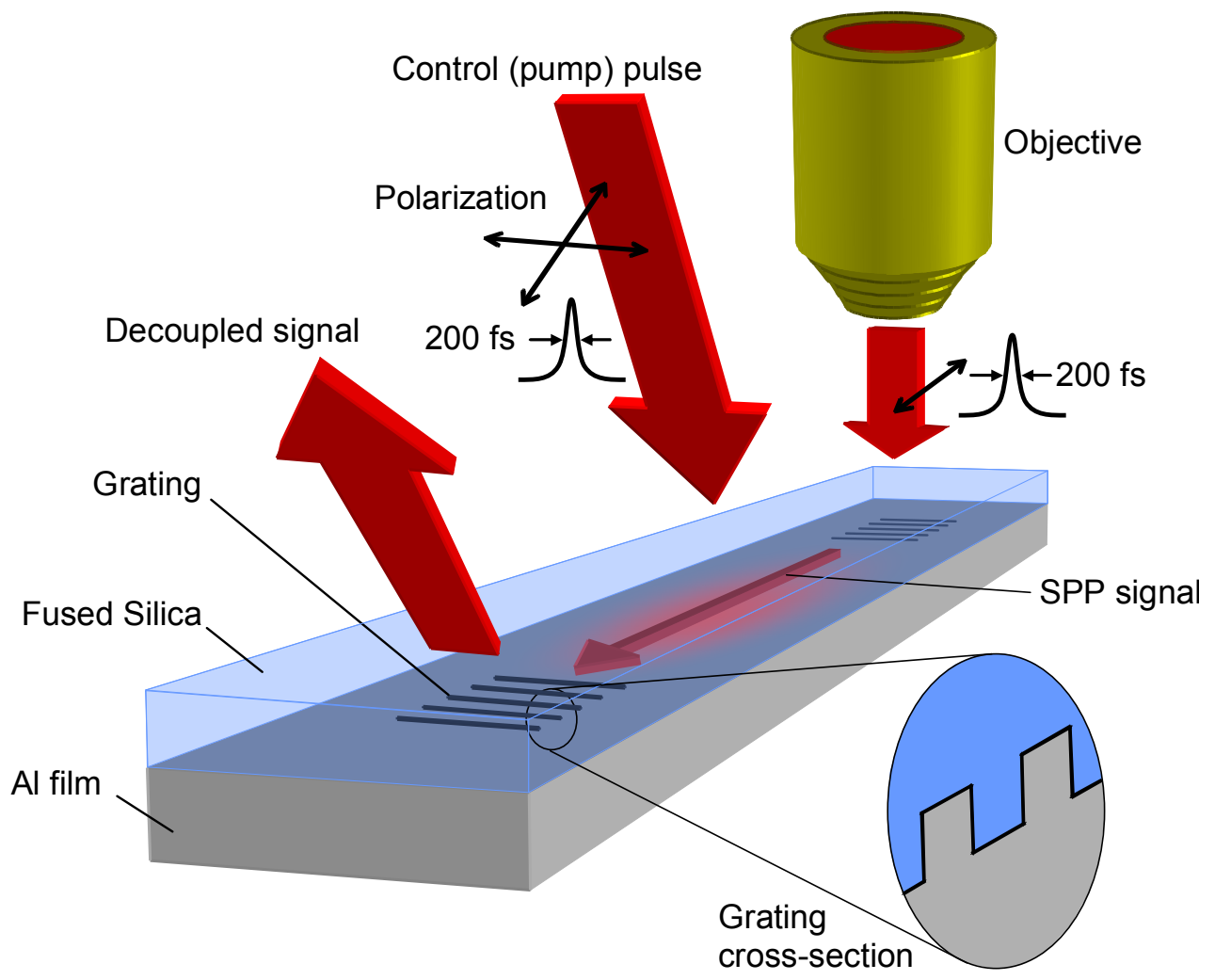

FIG 1: Ultrafast optical modulation of SPP propagation: A plasmonic signal, coupled to and from the waveguide by gratings on an aluminium/silica interface, is modulated by optical pump pulses as it travels between the gratings.

The waveguide structure and grating patterns were fabricated on optically polished fused silica substrates using electron beam lithography and anisotropic reactive ion etching. Grating structures were etched into the silica to a nominal depth of 43 nm while areas between the gratings were masked. The fabrication was completed with the evaporation of a 250 nm aluminium layer to form an optically flat metal/silica plasmon waveguide interface.

The plasmonic probe signal was generated on the Al/silica interface by grating coupling from a normally incident 780 nm pulsed laser beam and detected in the optical far field after decoupling at an oblique angle by a second grating separated from the first by 5 μm of unstructured metal/dielectric interface (a distance comparable to the SPP decay length). At 780 nm the photon energy of the optical radiation ($\hbar\omega$ = 1.59 eV) is close to the interband absorption peak in aluminium ($\hbar\omega$ = 1.55 eV), the metal component of the plasmon waveguide. With an electron configuration of [Ne]$3s^2 3p^1$ aluminium is a classic example of a nearly free-electron-like polyvalent metal. Its optical interband absorption originates mainly from transitions between parallel bands $\Sigma_3$ - $\Sigma_1$ in the vicinity of the $\Sigma$ [110] axis, near the $K$ point (see Fig. 2a and Ref. 8).

The coupling and decoupling gratings, each comprising 40 lines, had periods of 0.522 µm (optimized for normal incidence coupling) and 1.184 µm (giving an output beam angle of 54° after refraction at the silica/air interface) respectively. The beam from an amplified mode-locked Ti:sapphire laser (Coherent Mira + RegA) tuned to a centre wavelength of 780 nm, generating nearly transform-limited 200 fs optical pulses at a rate of 250 kHz, was split into pump and probe components, which were modulated at different frequencies ($\nu_1$ and $\nu_2$). The probe beam, polarized parallel to the grating vectors as required for coupling to a plasmon wave, was directed at normal incidence via a 10x long working distance objective, to a 17 µm diameter spot with a fluence of 0.9 mJ/cm$^2$ onto the coupling grating. The decoupled signal was monitored using a silicon photodetector and lock-in amplifier. The pump beam was focused onto the sample at an oblique angle (27°) to a spot with a diameter of 34 µm centred on the unstructured region between the coupling and decoupling gratings. An optical delay line was employed to vary the arrival time of pump pulses at the sample relative to the corresponding probe pulses, and the transient effect of pump excitation on the propagation of the probe SPP signal was monitored by recording the magnitude of the decoupled optical signal at the chopping sum frequency ($\nu_1$ + $\nu_2$) as a function of pump-probe delay. Group velocity dispersion for the plasmonic signal is close to zero around 780 nm and pulse broadening during propagation between the gratings is estimated to be no more than a few femtoseconds.

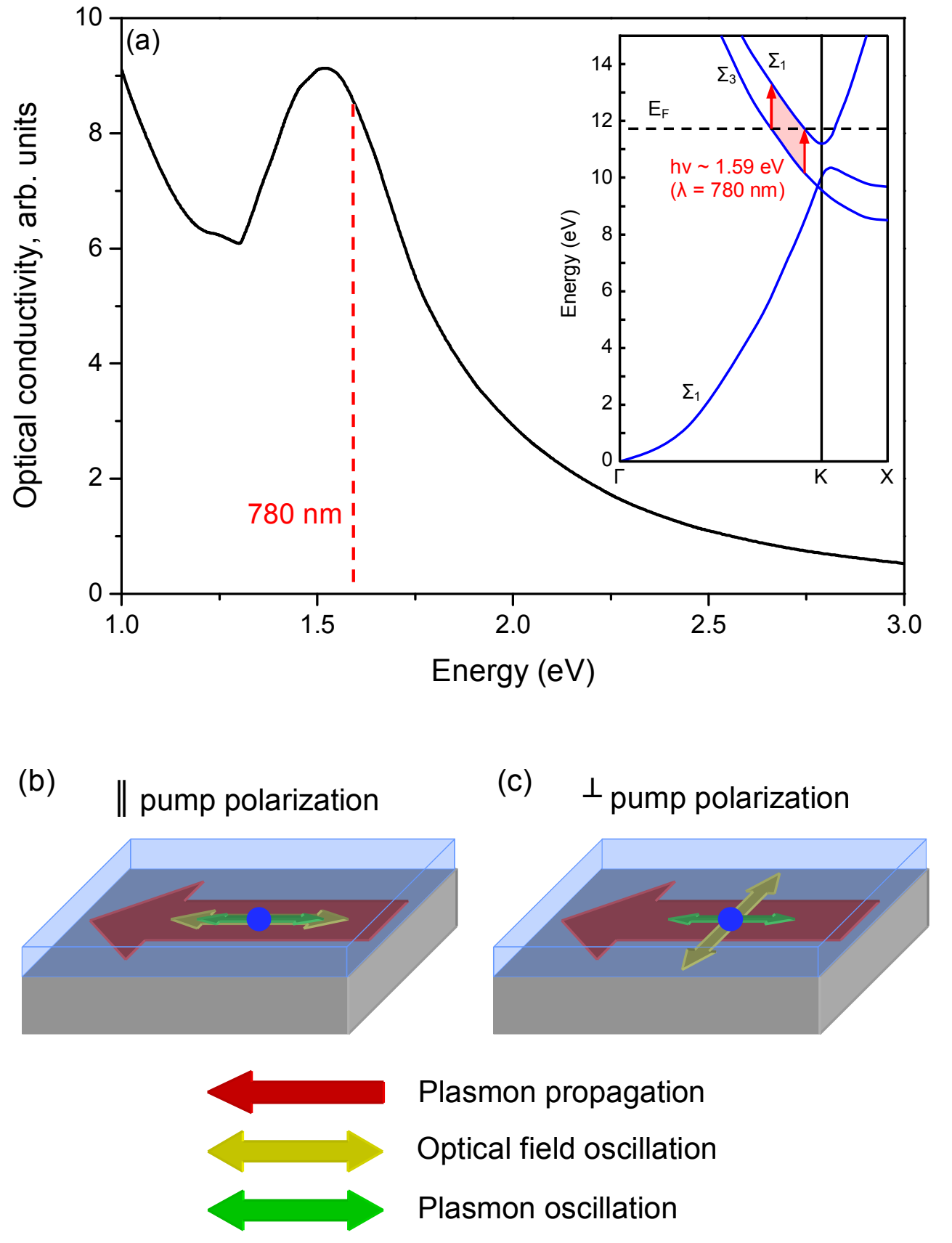

FIG 2: (a) Dispersion of the optical conductivity of aluminium showing the interband absorption peak. The inset shows the relevant part of the metal's band structure. (b) and (c) The two experimentally investigated configurations of pump pulse polarization and plasmonic electron oscillation directions.

Two experimental configurations were used: In the first, the linear polarization direction of the pump field was in the plane of incidence containing the SPP propagation direction and was thus predominantly in the direction of the electron oscillations in the SPP wave (see Fig. 2b). In the second configuration, the pump field polarization was perpendicular to the plane of incidence and was thus perpendicular to the electron oscillations in the plasmon wave (Fig. 2c).

The experimental results are summarized in Fig. 3, which shows the effect that the pump pulses have on the amplitude of the decoupled plasmonic signal as a function of pump-probe delay time (parts a and b) and pump fluence (part c). `Fast' and `slow' components of the transient pump-probe interaction have been observed. In all cases the presence of the pump pulse increases the magnitude of the transmitted plasmonic signal. The fast component replicates the optical cross-correlation function of the pump and probe pulses (Fig. 3a) but is only seen in the first experimental configuration, when the pump polarization has a component parallel to the direction of the SPP propagation. The magnitude of this fast SPP signal modulation component reaches a level of around 7.5% for pump pulse fluences of 10 mJ/cm$^2$ (Fig. 3c). The slow component of the transient response is present in both experimental configurations, i.e. for pump polarization directions both parallel and perpendicular to the SPP propagation direction. It grows for around 2 ps during and after a pump pulse then begins to relax with a characteristic decay time of about 60 ps (Fig. 3b). The magnitude of this slow SPP modulation reaches about 4% at a pump fluence of 10 mJ/cm$^2$ (Fig. 3c).

The transient response data indicate that there are two components to the nonlinear response: a fast component with a relaxation time shorter than the 200 fs pulse duration and a slow component with a relaxation time of about 60 ps. The fast component is sensitive to the mutual orientation of the pump beam's polarization state and the electron oscillation direction in the signal plasmon wave.

We believe that on the microscopic level the mechanisms underpinning the reported plasmon modulation effect are related to those responsible for transient changes in aluminium reflectivity observed in femtosecond pump-probe experiments[9,10]. Indeed, any pump-induced variation in refractive index $N = n + ik$ will simultaneously lead to changes in reflectivity $R = \frac{(1-n)^2 + k^2}{(1+n)^2 + k^2}$ and plasmon decay length $L = \frac{\lambda}{2\pi} \times \frac{(n^2 - k^2)^2}{2nk}$ given by the following formulae:

$$\delta R = \frac{4}{(n^2 + k^2 + 2n + 1)^2} \times \left[ (n^2 - k^2 - 1)\delta n + (2nk)\delta k \right]$$

$$\delta L = \frac{\lambda (n^2 - k^2)}{4\pi n^2 k^2} \times \left[ (3n^2 k + k^3)\delta n - (3nk^2 + n^3)\delta k \right]$$

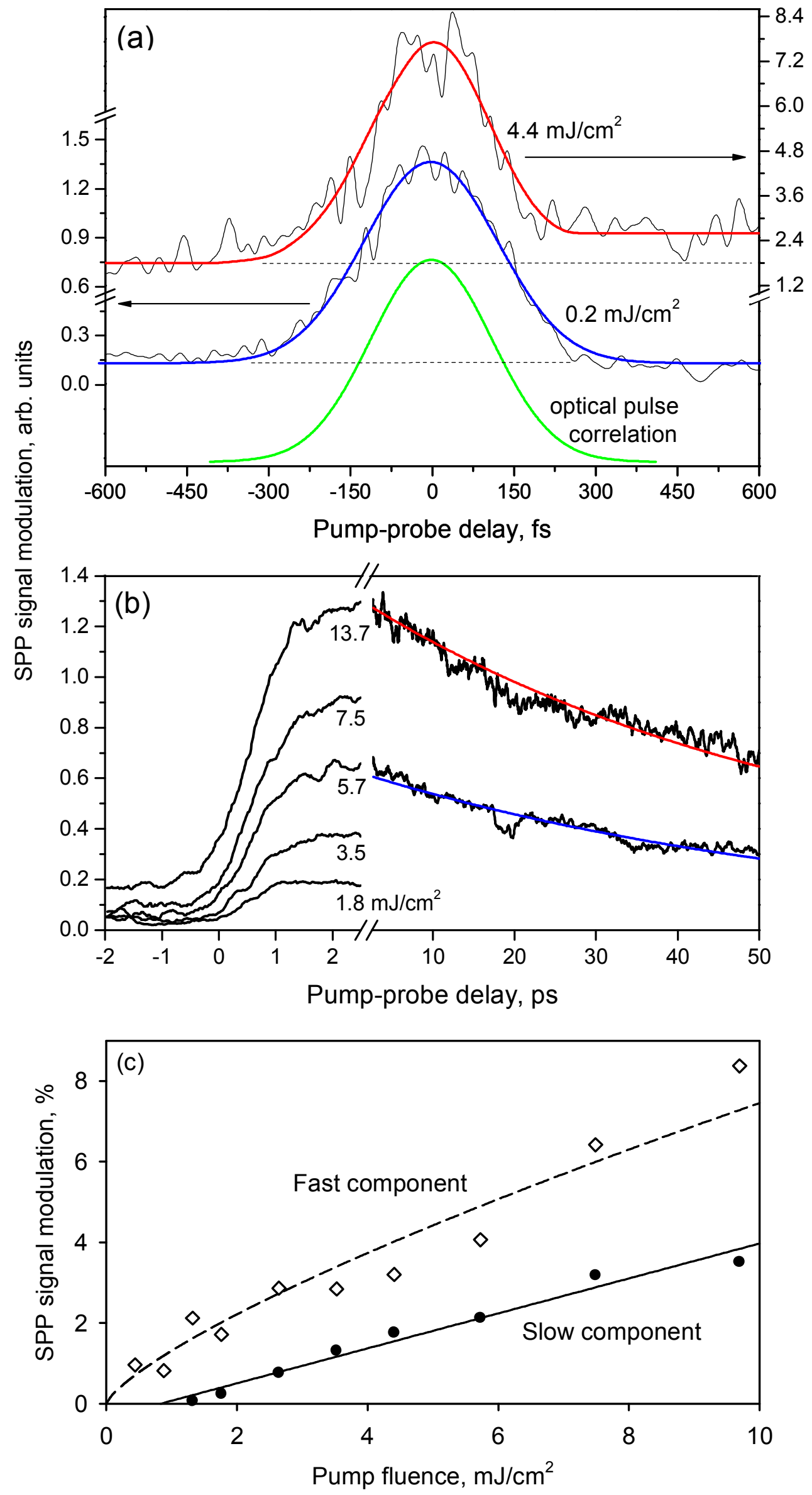

FIG 3: (a) Transient pump-induced changes in the decoupled plasmonic signal for pump light polarized parallel (a), and perpendicular (b) to the SPP propagation direction. (c) Corresponding peak magnitudes of the fast and slow pump-induced modulation components as a function of pump pulse fluence.

For metals it is generally true that $k > n$ and aluminium is no exception at the experimental wavelength $\lambda = 780$ nm. Under such conditions, $\frac{\partial L}{\partial n}$ and $\frac{\partial R}{\partial n}$ have the same sign, as do $\frac{\partial L}{\partial k}$ and $\frac{\partial R}{\partial k}$. Analysis shows that the pump-induced changes in aluminium's dielectric coefficients derived from reflectivity data in Ref. 9 would give rise to an increase in plasmon decay length as observed in the present study.

The connection between light-induced reflectivity increases and plasmonic signal propagation may also be illustrated using the Drude model, wherein it is found that a 7% change in the density of free carriers (induced by pump pulse excitation) gives an increase of 7.5% in the detected plasmon signal intensity (as observed at a fluence of 10 mJ/cm$^2$) by increasing the plasmon decay length. This increase in free carrier density simultaneously produces an increase of 0.48% in the reflectivity of the aluminium/silica interface. These figures are consistent with Guo *et al.*[9] who observed a reflectivity increase of ~0.8% at a fluence of 10 mJ/cm$^2$, and with Wilks and Hicken[10] who saw an increase of ~0.1% at about 1 mJ/cm$^2$.

A fast, polarization sensitive increase in aluminium reflectivity has previously been observed in optical pump-probe experiments[10], and as for the present case of light-SPP interaction it appeared only for parallel pump and probe polarizations. This zero-delay spike, which is also routinely seen in other metals, is due to a combination of a coherent nonlinearity and coupling of the wavelength degenerate pump into the probe via a transient grating created by the pump beam and probe SPP wave. The coherent nonlinearity is linked to anharmonic components of plasmonic oscillation resulting from the non-parabolicity of the electron dispersion. For high pump energies, the surface plasmon oscillation is strongly excited, giving rise to large amplitude, coherent collective electron oscillations. These lead to an increase in the damping of the SPP and the activation of additional nonlinear mechanisms[11]. The overall effect is to alter the short-time dynamics of the conduction-electron distribution and to reduce the efficiency of the coherent nonlinear interaction at higher fluences, giving rise to the observed sub-linear increase in the magnitude of the fast component with pump fluence (Fig. 3c). The disappearance of the fast component for perpendicular polarizations is characteristic of a nonlinearity related to the non-parabolicity of free-electron dispersion. It occurs for the same reasons that the third harmonic generated on reflection from a free-electron metal surface has the same polarization as the pump[12].

The slow component of the interaction between optical pump and SPP probe pulses shows no discernable dependence on the mutual orientation of pump and probe polarizations. It has the same origin as the slow transient reflectivity change observed in aluminium[9]: When an intense pump pulse excites numerous electrons to states above the Fermi level through an interband transition, a transient response known as the 'Fermi smearing' nonlinearity occurs rapidly then disappears as electrons thermalize with the lattice. Subsequent changes in the dielectric coefficients are essentially of a thermal and elastic nature. In this case the relaxation time of the response is related to the time needed for heat to leave the metal's skin layer and for the lattice deformation to recover. We argue that the increasing magnitude of the slow response component observed during the first two picoseconds after excitation (see Fig. 3b) is related to a dynamic balance between Fermi smearing and thermal/elastic effects, while the observed relaxation time of ~60 ps is related to the thermal and elastic transients in the skin layer and is in full agreement with Ref. 13.

It should be noted that the level of direct modulation of the plasmonic signal may be increased manyfold using interferometric arrangements, as has been demonstrated with plasmonic versions of Mach-Zender[4] and Fabry-Perot[7] interferometers. The demonstrated switching time is around 200 fs but could be as short as a few tens of femtoseconds as it is ultimately limited only by the electron momentum relaxation time. This is radically faster than the millisecond response time of thermo-plasmonic modulators[4,5]. The required excitation level compares favourably with the optical fluence of about 15 mJ/cm$^2$ required to control a plasmonic gate based on structural phase switching in a gallium plasmon waveguide, where switching times of about 50 ns are achieved[14]. The switching fluence for an optical pulse acting directly on the aluminium waveguide is a factor of a hundred higher than required in a plasmonic modulator exploiting CdSe quantum dots to control SPP losses, but the 200 fs switching time achieved in the present study is more than five orders of magnitude shorter than the 40 ns switching time reported for the quantum dot device[7].

In summary, we report the first experimental evidence that femtosecond plasmon pulses can be generated, transmitted, decoupled and detected, and describe a new principle for the direct optical modulation of plasmon signals with terahertz bandwidth that is supported by experimental demonstration.

This work was supported by the Engineering and Physical Sciences Research Council (UK), the Department of Energy and National Science Foundation (USA), and the United States-Israel Binational Science Foundation. The authors would like to acknowledge the technical assistance of J. D. Mills.

---

* niz@orc.soton.ac.uk; www.nanophotonics.org.uk